\documentclass[aps,twocolumn,prl]{revtex4}
\usepackage{graphicx}

\newcommand{\beq}{\begin{equation}}
\newcommand{\eeq}{\end{equation}}
\newcommand{\beqa}{\begin{eqnarray}}
\newcommand{\eeqa}{\end{eqnarray}}

\def\eGS{e_{\rm GS}}
\def\eLQ{e_{\rm LQ}}
\def\eCR{e_{\rm CR}}
\def\fLQ{f_{\rm LQ}}
\def\fCR{f_{\rm CR}}

\begin{document}

\title{Glassy dynamics, metastability limit and crystal growth in a
lattice spin model} 

\author{Andrea Cavagna, Irene Giardina, Tomas Grigera}

\affiliation{Center for Statistical Mechanics and Complexity, INFM Roma ``La Sapienza''}  
\affiliation{Dipartimento di Fisica, Universit\`a di Roma ``La Sapienza'', 00185 Roma, Italy}
             
\date{July 5, 2002}

\begin{abstract}
We introduce a lattice spin model where frustration is due to
multibody interactions rather than quenched disorder in the
Hamiltonian.  The system has a crystalline ground state and below the
melting temperature displays a dynamic behaviour typical of fragile
glasses. However, the supercooled phase loses stability at an
effective spinodal temperature, and thanks to this the Kauzmann
paradox is resolved. Below the spinodal the system enters an
off-equilibrium regime corresponding to fast crystal nucleation
followed by slow activated crystal growth.  In this phase and in a
time region which is longer the lower the temperature we observe a
violation of the fluctuation-dissipation theo\-rem analogous to
structural glasses. Moreover, we show that in this system there is no
qualitative difference between a locally stable glassy configuration
and a highly disordered polycrystal.
\end{abstract}

\maketitle

When cooled fast enough, most liquids can be brought below their
melting temperature $T_m$ while avoiding crystallization.  In this
supercooled phase the viscosity increases with decreasing temperature,
until at the glass transition $T_g$ relaxation time becomes too long
and equilibration cannot be achieved within experimentally accessible
times.  In 1948, W.~Kauzmann \cite{kauzmann} noted that if the entropy
of a supercooled liquid is extrapolated below $T_g$, it becomes equal
to the crystal entropy at a temperature $T_s>0$, and even smaller than
zero if extrapolated further. This {\it entropy crisis} is never
actually observed, because the glass transition intervenes
before. However, Kauzmann found it paradoxical that it was just
a kinetic phenomenon (the glass transition) that saved the liquid from
a thermodynamic nonsense.

In the context of the Adams-Gibbs  theory \cite{adams}, the
entropy crisis has however an interpretation: the entropy difference
between crystal and liquid is related to the configurational entropy
$\Sigma$, that is the entropic contribution due to the presence of an
exponentially high number of different glassy minima. The vanishing of
$\Sigma$ at $T_s$ signals a thermodynamic transition to a new phase,
characterized by a sub-exponential number of glassy states, separated
by infinite free-energy barriers. This picture is exact for some
mean-field spin-glass systems \cite{pspin}, and it may be the correct
resolution of the Kauzmann paradox even for real structural glasses.

Despite analytic and numerical work 
supporting the entropy crisis scenario \cite{evidence}, there is another way to
avoid the Kauzmann paradox, which, interestingly enough, was proposed
by Kauzmann himself \cite{kauzmann}.  He rejected the idea of a
thermodynamic glassy phase, and of a transition at $T_s$. What
Kauzmann hypothesized is the existence of a metastability limit of the
supercooled liquid phase, below which crystal nucleation becomes
faster than liquid equilibration.  More precisely, he defined an
effective spi\-no\-dal temperature $T_{\rm sp}>T_s$ below which {\it ``the free
energy barrier to crystal nucleation becomes reduced to the same height
as the barrier to simpler motions''}.  Below $T_{\rm sp}$ the supercooled
liquid is operationally meaningless and thus the paradox is
avoided. However, the metastability limit may prove impossible to
observe experimentally if the equilibration time at $T_{\rm sp}$ is
much larger than the experimental time, that is if $T_{\rm sp}<T_g$.

In the Kauzmann scenario, the off-equilibrium phase below $T_{\rm sp}$
basically consists in a very slow crystal domain growth.  This
suggests two criteria to detect whether a system has such a
metastability limit or not, even when $T_{\rm sp}$ is experimentally
inaccessible.  First, we may think that there is a {\it qualitative}
difference between a disordered glassy configuration, obtained by
quenching a liquid, and a polycrystalline configuration, however rich
in defects this is, and however slow crystal growth may be.  Secondly,
we know that the off-equilibrium dynamics of simple domain growth (as
in the Ising model) can be distinguished from glassy dynamics (as
observed in structural glasses) by a different violation of the
fluctuation-dissipation theorem (FDT) \cite{cuku-peliti,x-vetri}.

In this Letter we present a model where the Kauzmann paradox is
avoided by a metastability limit. The model is thus a good test
for the two criteria discussed above. We shall find that neither criterion 
is sharp 
enough to discriminate such a system from a typical structural glass.
More precisely, if our experimental time were not long enough to {\it
explicitly} observe the loss of stability of the liquid at $T_{\rm sp}$, it
would be impossible to distinguish the present system from an ordinary
fragile glass. The reason is that below $T_{\rm sp}$ crystal {\it
nucleation} is fast, but crystal {\it growth} becomes very 
slow, with many crystal droplets trying to expand in a liquid
background \cite{glacial}. In such a situation 
distinguishing between a truly disordered glass and a mixture of tiny
mismatched crystallites becomes very hard, and FDT violation 
is nontrivial.

The aim of our study is twofold.  First, we want to show that
Kauzmann's resolution of the Kauzmann paradox is valid at least in one
simple system, and that it is not necessarily in conflict with glassy
phenomenology.  Second, we hope this example will help to develop some
more strict criteria to distinguish systems with a thermodynamic
transition at $T_s$, from those where the Kauzmann scenario holds.  
The Hamiltonian of our model is,
\begin{equation}
H=\sum_{i=1}^N (1+s_i) f_i, \qquad
	f_i=s_i^W s_i^S s_i^E s_i^N ,  \label{ctls}
\end{equation}
where $W$ is for west, $S$ for south, etc. The spins $s_i=\pm 1$
belong to a two-dimensional square lattice of linear size $L$. 
We perform single-spin-flip Monte Carlo simulations in square lattices
with $L=100$ and $L=500$ \cite{size}.  

The disordered version of this model was
first introduced in \cite{tomas-ctls} to describe an ensemble of coupled
two level systems, whence its name, CTLS. 
There has been recently much interest in lattice models of this sort, where
multibody interactions in the Hamiltonian ensure frustration even without 
quenched disorder \cite{shore,lipo,juanpe-plaq,meza}.

The CTLS has a crystalline ground state
obtained by covering the lattice with the following non-overlapping
$5$-spins elements: $s_i=-1, s_i^{W}=s_i^{S}= s_i^{E}=s_i^{N}=+1$. The
ground state energy density is $\eGS=-1.6$.  The unit cell size is
$5\times 5$, and this together with the symmetry $x\to \mp x, y\to \pm
y$, gives a ground state degeneracy of $50$. The finite degeneracy of
the crystal is a key feature of the CTLS compared to the plaquette
model of \cite{lipo}, since it will allow us to directly measure the amount 
of crystalline order in the system.

Model (\ref{ctls}) has a first order (melting) transition.  To locate
the melting temperature $T_m$ we compute the free energy $f$ for the
crystal (CR) and liquid (LQ) and ask that $\fCR(T_m)=\fLQ(T_m)$. The
free energy is obtained integrating the energy ($\beta=1/T$): $\beta
f(\beta)=\beta_0 f(\beta_0) + \int_{\beta_0}^\beta \!\! d\beta' \,
e(\beta')$, taking $\beta_0=0$ for the liquid, and $\beta_0=\infty$
for the crystal.  The equilibrium energies are well fitted by,
\begin{eqnarray}
\eLQ(T) &=& -1.8\, \tanh(1.1/T)\ ,
\label{energy}  \\
\eCR(T) &=& \eGS + 5.5\times 10^{-5}\ T^{13} \ .
\nonumber
\end{eqnarray}
Using these relations we find $T_m=1.30$. 
Unless cooling is exceedingly slow, crystallization is not attained
at $T_m$, and the liquid can be kept at equilibrium in its supercooled phase
$T<T_m$.
Extrapolation of $\eLQ(T)$ and $\fLQ(T)$ gives the temperature $T_s$ where 
the entropy of the supercooled liquid equalizes that of the crystal, 
namely the Kauzmann paradox temperature. We find $T_s=0.91$.

\begin{figure}
\includegraphics[clip,width=\columnwidth]{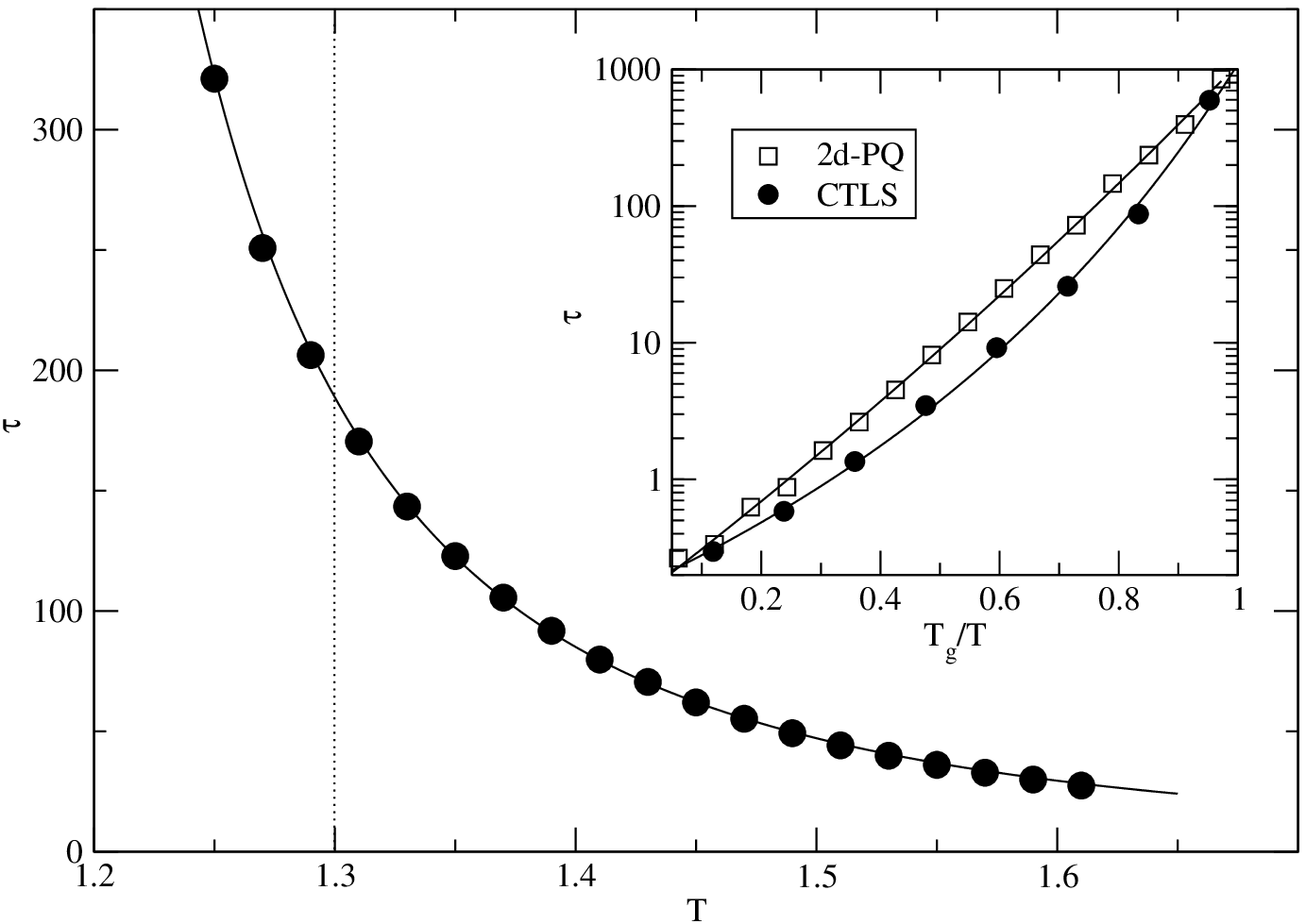}
\caption{Relaxation time as a function of the temperature. Full line:
power law fit.  Inset: fragility plot for a comparison of the CTLS
with the model studied in \cite{juanpe-plaq}. $T_g$ is an operational 'glass transition', 
defined by $\tau(T_g)=1000$. $L=500$.}
\label{MINI-AA}
\end{figure}

The equilibrium dynamics of the CTLS in the supercooled
phase can be studied by measuring the normalized spin-spin 
correlation function, $ C_n(t,t_w)=[ \langle s(t_w) s(t) \rangle - \langle
s\rangle^2]/[1-\langle s\rangle^2]$, with $t>t_w$.  In equilibrium 
$C_n(t,t_w)=C_n(t-t_w)$, and we find that the correlation can be fitted to
a stretched exponential, $C_n=\exp\left[-(t-t_w/\tau)^\beta\right]$.  In
Fig.~\ref{MINI-AA} we plot the relaxation time $\tau$ as a function of
$T$, together with a power law fit $\tau=A/(T-T_c)^\gamma$, with
$T_c=1.06$, $\gamma=2.29$. The accuracy of the fit suggests that the
CTLS is a fragile system, a fact supported by the Angell plot \cite{angell2} in the 
inset of Fig.~\ref{MINI-AA}, which compares the relaxation times of the CTLS
and a strong system, the two-dimensional plaquette model (2$d$-PQ) studied in
\cite{juanpe-plaq}. Data are fitted to a Vogel-Fulcher
form, $\tau=\tau_0 \exp[\Delta/(T-T_0)]$. In the CTLS we find $T_0=0.76$ 
for $T\in[1.2:1.7]$, and $T_0=0.90$ for $T\in[1.2:1.4]$.

\begin{figure}
\includegraphics[clip,width=\columnwidth]{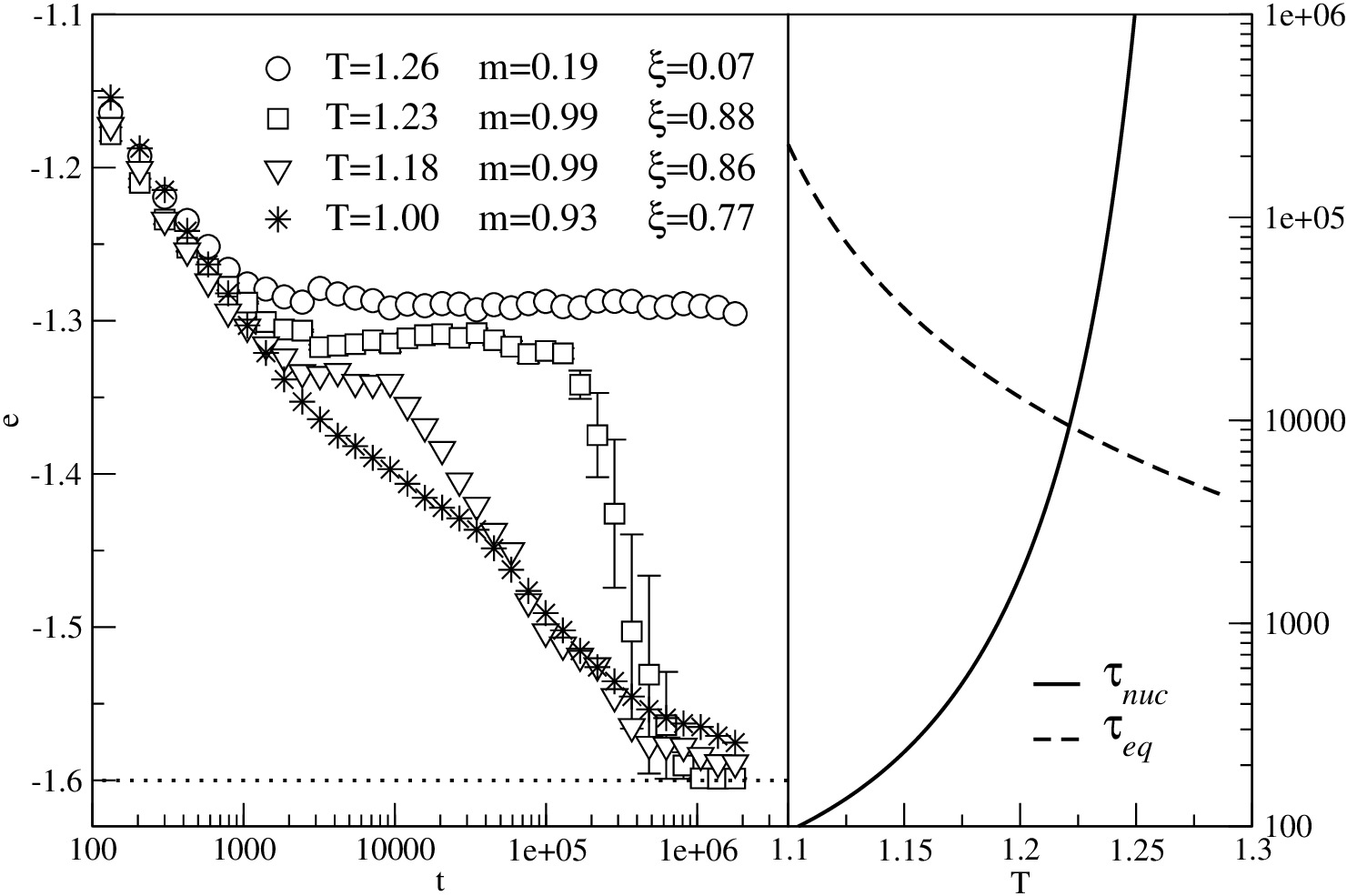}
\caption{Left: $e$ vs.\ $t$ at different temperatures, $L=100$. Error bars are
only showed when larger than symbols size. The values of $m$ and $\xi$
in the legend are measured at the longest time for each temperature (see text). 
Right: crystal
nucleation time and liquid equilibration time vs.\ temperature.}
\label{MINI-BA}
\end{figure}

Below about $T\approx 1.2$ the relaxation time cannot be
measured, because crystal nucleation starts before the 
li\-quid equilibrates.
This can be seen in
Fig.~\ref{MINI-BA}, where we plot the energy density vs.\ time at four
different temperatures below $T_m$, with random initial condition.  At
$T=1.26$, the system relaxes in the supercooled liquid and remains in
this phase up to our experimental time, $2\times 10^6$ MCS.  On the
other hand, for $T=1.23$, after initial equilibration in the liquid phase,
the system makes a sharp transition to the crystal: on average crystal nucleation 
starts at about $10^5$ MCS, while crystal growth is completed in $10^6$ MCS.  At
$T=1.18$, however, nucleation starts roughly after $10^4$ MCS,
while complete crystallization is achieved in more than $10^6$ MCS. 
At $T=1$, the liquid plateau disappears and crystallization is
still incomplete after the largest time. Thus at lower temperatures nucleation is 
faster, but crystal growth is slower. 

To pinpoint the liquid metastability limit we use standard nucleation 
theory \cite{nuc}.
A necessary condition for the existence of the supercooled liquid is
that the crystal nucleation time $\tau_{nuc}$ is much longer than the liquid equilibration
time $\tau_{eq}$.  We can estimate $\tau_{nuc}(T)$
close to a reference temperature $T^\star$, assuming
that the surface tension is constant. We have,
\begin{equation}
\tau_{nuc}(T)= \, \exp 
\left\{\frac{T^\star \, \delta f(T^\star)}{T \, \delta f(T)} \,
\log\left[\tau_{nuc}(T^\star)\right]\right\} \ ,
\end{equation}
where $\delta f(T)=\fLQ(T)-\fCR(T)$ is the bulk free energy difference
between supercooled liquid and crystal. We have chosen
$T^\star=1.23$.  To estimate the liquid equilibration time we note
that the correlation $C(t)$ drops to zero in about $20$ relaxation
times, and thus $\tau_{eq}=20 \,\tau$. By imposing $\tau_{nuc}(T_{\rm sp}) =
\tau_{eq}(T_{\rm sp})$, we obtain the effective spinodal temperature $T_{\rm sp}$
marking the metastability limit.  Fig.~\ref{MINI-BA} (right) shows
that $T_{\rm sp}=1.22$ \cite{bray}. We have $T_{\rm sp}>T_s$, and the Kauzmann 
paradox is thus avoided.

\begin{figure}
\includegraphics[clip,width=\columnwidth]{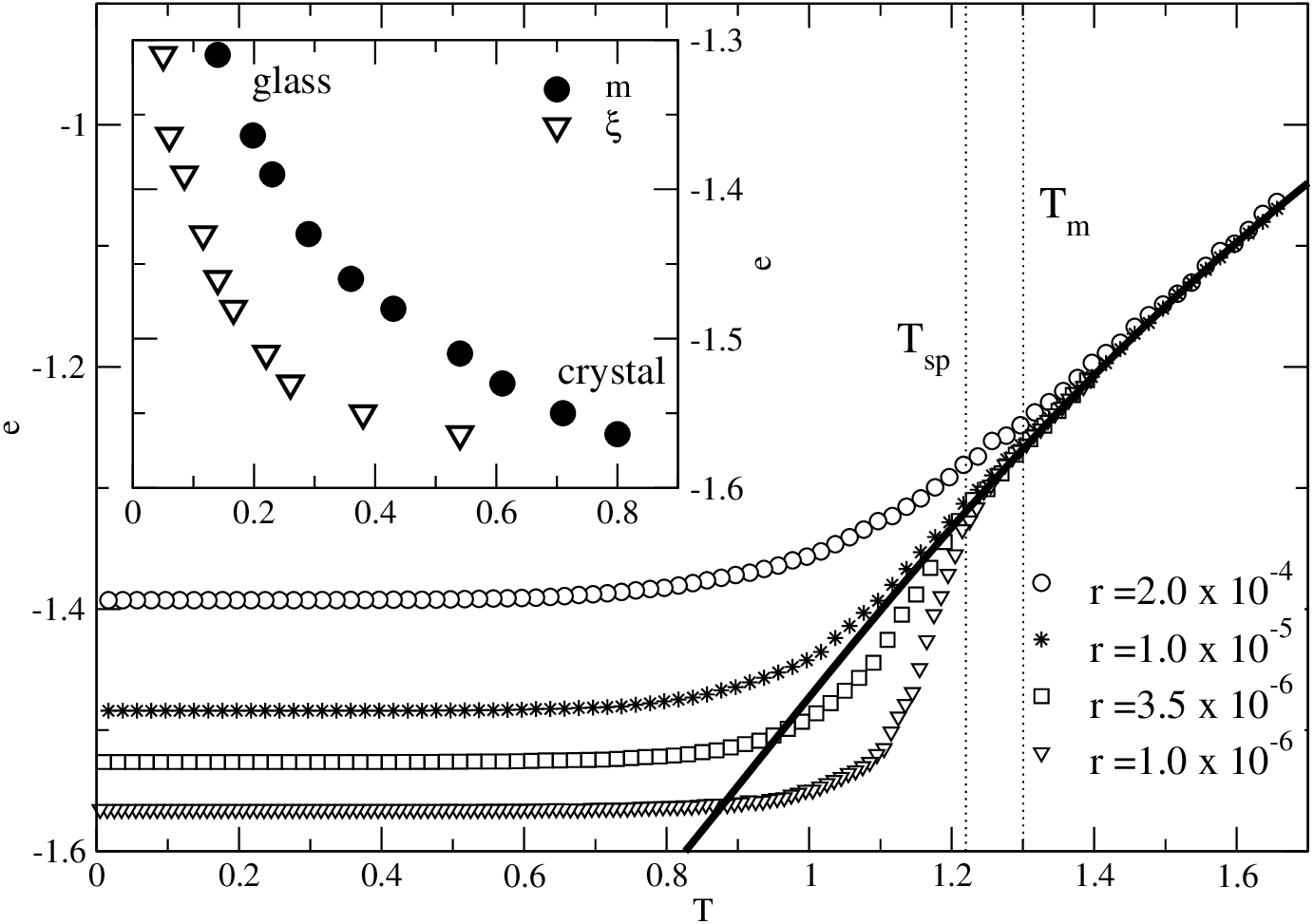}
\caption{$e$ as a function of $T$, at various cooling rates $r$. 
Full line: equilibrium liquid energy, eq. (\ref{energy}).
Inset: $e$ vs $m$ and $\xi$ in the $T=0$ asymptotic states
($r\in[10^{-6},10^{-3}]$). $L=100$.}
\label{MINI-DA}
\end{figure}

Below the metastability limit $T_{\rm sp}$ the only
equilibrium phase is the crystal. However, a long-lived 
off-equilibrium glassy phase can still be formed:
if we cool the system, it eventually remains stuck in an 
off-equilibrium state whose asymptotic energy is lower the
slower the cooling rate $r$ (Fig.~\ref{MINI-DA}). 
All the configurations reached at $T=0$ are stable minima, 
indicating that activation is needed to grow the crystal. 
For very fast 
coolings these configurations are completely disordered, while at the 
slowest cooling rates they correspond to highly ordered polycrystals. 
Is there a {\it qualitative} difference among these asymptotic 
states ? In other words, is it possible to sharply separate a {\it bona fide} 
glass from the polycrystal ? 

To answer this question we introduce two direct measures of
crystalline order. The first one is the normalized domain size $\xi$,
obtained from the Fourier transform $G(k)$ of $G(r) = \langle s_i
s_{i+r} \rangle$.  $G(k)$ has a peak at $k_0=2\pi/5$, and we define
$\xi$ as the inverse of the peak width, normalized by $L$. The second
is the normalized crystal mass $m$, i.e. the total fraction of
crystallized spins (defined as the number of spins down surrounded by
8 spins up, multiplied by $5$) divided by $L^2$ \cite{zero}.  
Using $\xi$ and $m$ we see that 
crystallization has been achieved at the lowest three temperatures of
Fig.~\ref{MINI-BA}. If we now plot the energy of the $T=0$ asymptotic 
configurations in a cooling experiment vs their crystalline mass and 
domain size, (Fig.~\ref{MINI-DA}, inset), we find a continuous spectrum of 
states.  Thus, in the CTLS there is structural continuity between highly 
disordered glassy minima and strongly ordered polycrystalline minima.
The answer to the question above is therefore negative.  
Yet, had our slowest cooling rate been $r=2\times 10^{-4}$ we would only observe 
a disordered glass.

\begin{figure}
\includegraphics[clip,width=\columnwidth]{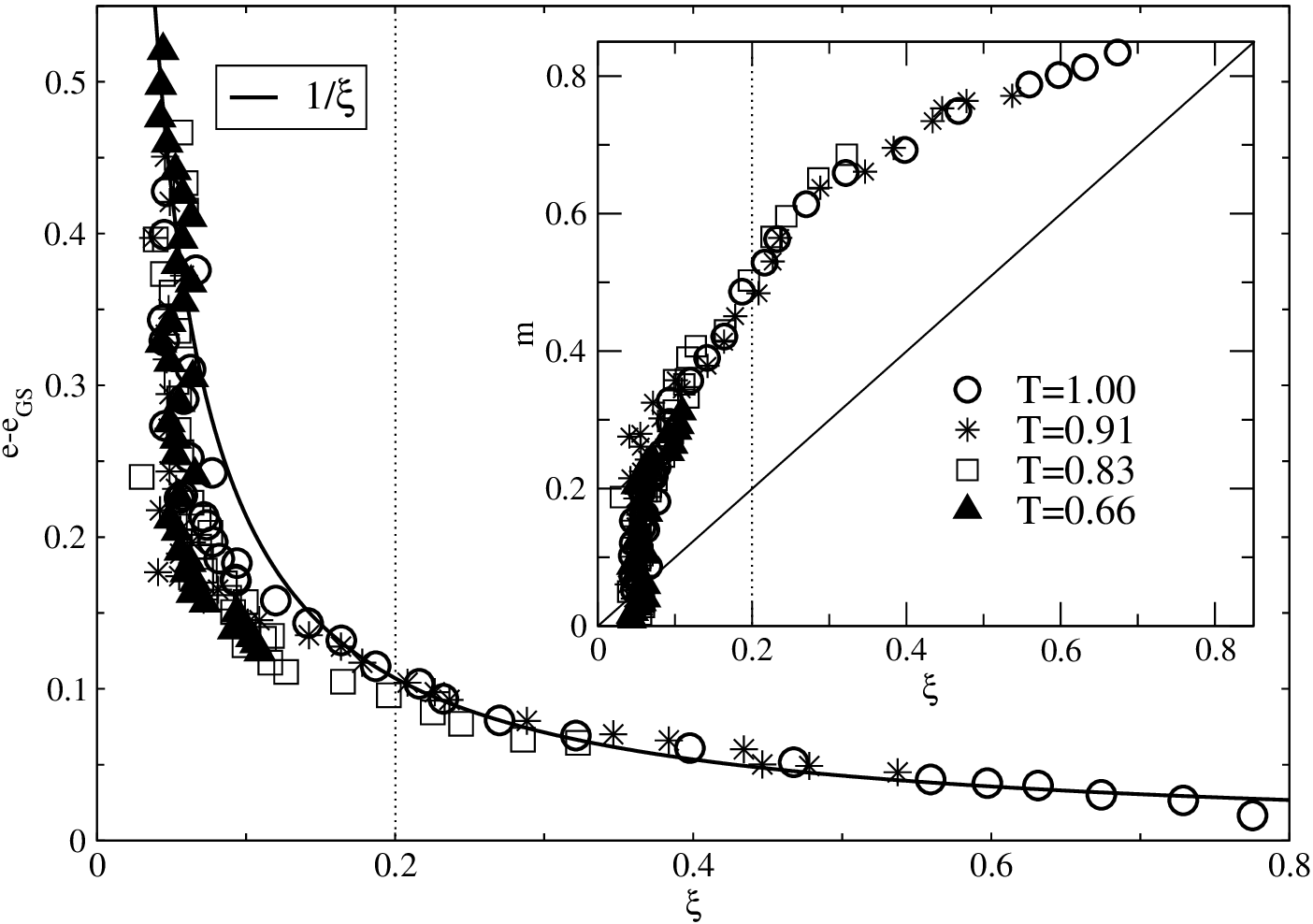}
\caption{
Excess energy $e-\eGS$ as a function of $\xi$. Full line: $1/\xi$ fit
of the data. 
Inset: $m(t)$ as a function of $\xi(t)$, parametrically in $t$,
after a quench at various temperatures. All runs are $2\cdot 10^6$ MCS
long. $L=100$.}
\label{MINI-EA}
\end{figure}

If not by structural difference, we might expect the dynamic behaviour
of our ``glass'' to betray its nature of crystal-growth phase, by displaying
a characteristic {\em coarsening} dynamics. In coarsening, excess
energy over the ground state is concentrated in the interfaces among
domains. This gives $e(t)-\eGS \propto \xi(t)^{d-1}/\xi(t)^d
= 1/\xi(t)$, for $L \xi \gg a$, where $a$ is
the interfacial width. We do in fact find such a regime 
(Fig.~\ref{MINI-EA}) for $\xi>0.2$. However, due to slow activated 
dynamics, the early regime, $L \xi\sim a$, can last for long, and in fact it is 
the only one that can be observed at the lowest temperatures. This regime, which we call
{\em bubbling,} is characterized by a rapid increase of $m$ at roughly
zero $\xi$ (Fig.~\ref{MINI-EA}, inset), meaning that fast nucleation
leads to rapid formation of many tiny crystal droplets. Only at
longer times, when most of the system has crystallized, the domains 
grow at expense of each other and proper coarsening starts. 
For large systems we expect $m \sim 1- a / L \xi$, such that bubbling and 
coarsening regimes become well separated.  
Summarizing, at low temperatures, or for short experimental times, the 
coarsening regime is inaccessible, in the same way as, for fast cooling, 
polycrystals are not observed.

\begin{figure}
\includegraphics[clip,width=\columnwidth]{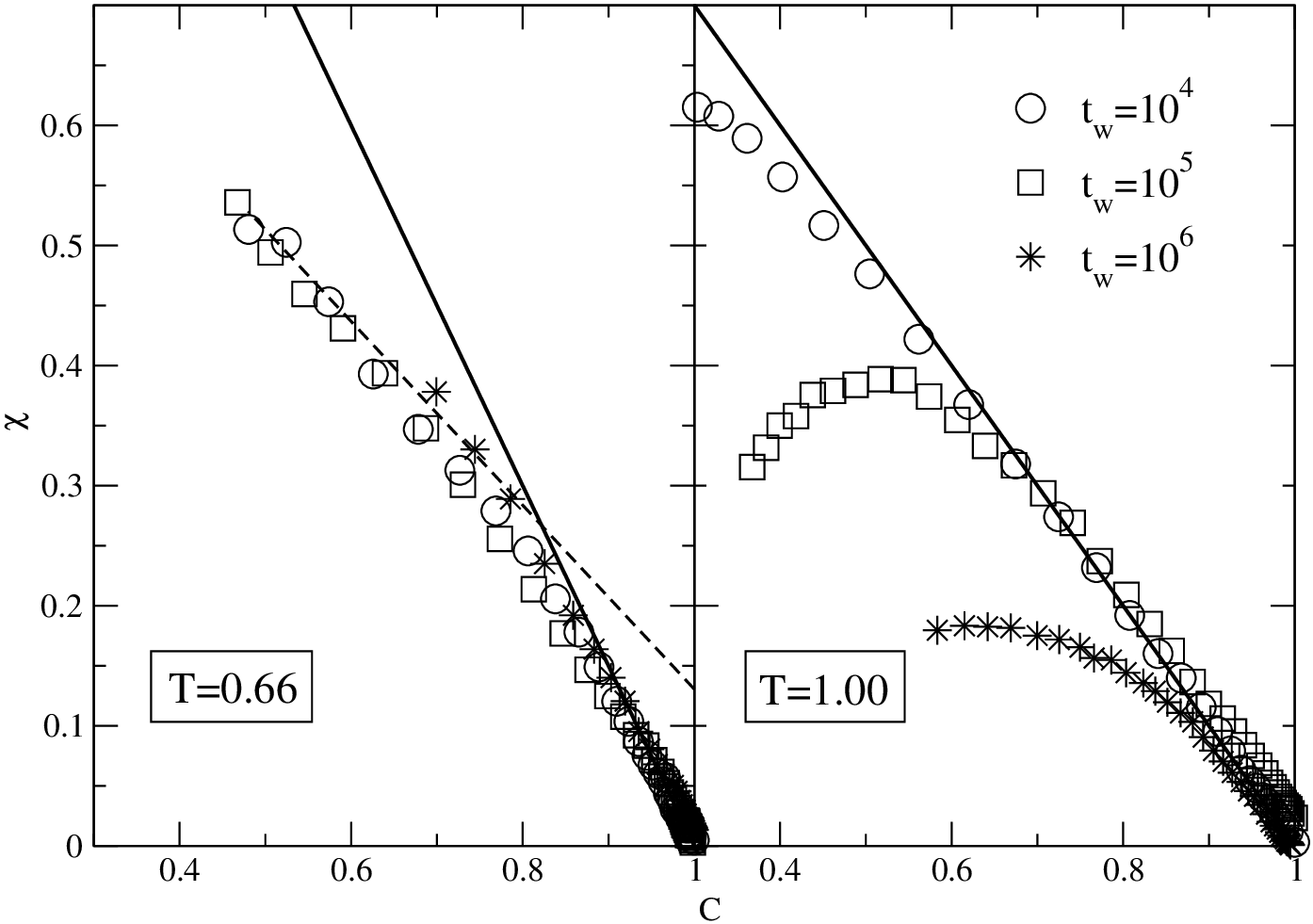}
\caption{Parametric plot of $\chi(t,t_w)$ vs 
$C(t,t_w)$, at various values of the waiting time $t_w$.  
Full lines represent the
equilibrium relation $\chi=\beta(1-C)$. For $T=0.66$ the broken line
is a linear fit for $C<0.8$. $L=500$.}
\label{MINI-FA}
\end{figure}

This behaviour has important consequences on the FDT violation
pattern. 
The integrated response is given by $\chi(t,t_w)=\int_{t_w}^t dt'\, R(t,t')$,
where $R(t,t')=\delta \langle s(t)\rangle/\delta h(t')$ and $h$ is the conjugate
field to the spin. We can make a parametric plot
of $\chi(t,t_w)$ vs the correlation $C(t,t_w)=\langle s(t) s(t_w) \rangle$. 
At equilibrium FDT holds and $\chi=\beta(1-C)$, while a 
departure from FDT is normally observed
for late times if the system is out of equilibrium.  This FDT
violation is the standard tool to discriminate genuine glassy
behaviour from simple domain growth. The relation $\chi=\beta X (1-C)$ 
can be used to express the late time FDT violation in off-equilibrium systems.
For simple domain growth $X=0$, while in structural glasses it is
observed $0<X<1$ \cite{cuku-peliti,x-vetri,early}.  
We do expect a trivial FDT violation ($X=0$) in the coarsening regime, 
but the situation may be more complicated in the earlier bubbling regime.
From Fig.~\ref{MINI-EA} we see that at $T=0.66$ the system remains in
the bubbling regime up to the largest time. At the same temperature
we find an FDT violation very similar to structural glasses
(Fig.~\ref{MINI-FA}): for $C$ smaller than a given breaking
point the slope changes, giving rise to a constant nonzero value of
$X$. Moreover, the pattern basically does not change within two orders
of magnitude in the waiting time $t_w$.  At $T=0.66$ we can thus
define an effective temperature $T_{eff}=T/X=1.32$.
On the other hand, for $T=1$ as $t_w$ increases we see a
crossover from a nontrivial FDT violation ($X\neq 0$), to a seemingly
trivial pattern ($X\sim0$). This is consistent with the fact that 
at $T=1$ the coarsening regime is reached within our
experimental time (Fig.~\ref{MINI-EA}).  Yet, once again, had our longest 
experimental time been too short to enter the coarsening regime, we would 
only observe a glassy FDT violation.

In this Letter we presented a model where the metastability limit
can be either observed directly as a loss of stability of the 
supercooled liquid, or indirectly, thanks to the formation of
polycrystalline asymptotic states in slow cooling experiments, and
thanks to trivial FDT violation in the late coarsening regime. 
However, in order to be effective the indirect methods 
need as long an experimental time as the direct method. 
For shorter time scales the present model is compatible 
with ordinary fragile glasses.

We acknowledge many important discussions with P.~Debenedetti, H.~Horner,
V.~Martin-Mayor, G.~Parisi, F.~Ricci-Tersenghi, F.~Sciortino, D.~Sherrington
and P.~Verrocchio. AC thanks in particular M.~A.~Moore.


\begin{thebibliography}{22}
\bibitem{size} Data obtained at $L=500$ and $L=100$ are quantitatively consistent.
\bibitem{zero} Both $\xi$ and $m$
have an effective zero given by their value in a random configuration,
$\xi_0=0.05$, $m_0=0.01$.
\end{thebibliography}
\end{document}